\documentclass[12pt]{article}

\usepackage{amsmath}
\usepackage{amssymb}
\usepackage{amsfonts}
\usepackage{latexsym}
\usepackage{color}

\catcode `\@=11 \@addtoreset{equation}{section}

\catcode `\@=12

\newcommand{\be}{\begin{equation}}
\newcommand{\en}{\end{equation}}
\newcommand{\bea}{\begin{eqnarray}}
\newcommand{\ena}{\end{eqnarray}}
\newcommand{\beano}{\begin{eqnarray*}}
\newcommand{\enano}{\end{eqnarray*}}
\newcommand{\bee}{\begin{enumerate}}
\newcommand{\ene}{\end{enumerate}}

\newcommand{\N}{\mathfrak N}

\newcommand{\mc}{\mathcal}

\newcommand{\D}{{\mc D}}

\newcommand{\F}{{\cal F}}
\newcommand{\G}{{\cal G}}

\newcommand{\PP}{\mc P}
\newcommand{\Lc}{{\cal L}}

\newcommand{\1}{1 \!\! 1}

\newcommand{\Hil}{\mc H}

\catcode `\@=11 \@addtoreset{equation}{section}
\catcode `\@=12

\begin{document}

\thispagestyle{empty}

\vspace*{2cm}

\begin{center}
{\Large \bf Examples of Pseudo-bosons in quantum mechanics}   \vspace{2cm}\\

{\large F. Bagarello}\\
  Dipartimento di Metodi e Modelli Matematici,
Facolt\`a di Ingegneria,\\ Universit\`a di Palermo, I-90128  Palermo, Italy\\
e-mail: bagarell@unipa.it

\end{center}

\vspace*{2cm}

\begin{abstract}
\noindent We discuss two physical examples of the so-called {\em pseudo-bosons}, recently introduced in connection with pseudo-hermitian quantum mechanics. In particular, we show that the so-called {\em extended harmonic oscillator} and the {\em Swanson model} satisfy all the assumptions of the pseudo-bosonic framework  introduced by the author. We also prove that the biorthogonal bases they produce are not Riesz bases.

\end{abstract}

\vspace{2cm}


\vfill


\newpage

\section{Introduction}

In a series of recent papers, \cite{bag1}-\cite{bag4}, a family of {\em excitations} generalizing bosons have been introduced by considering an extended version of the canonical commutation relation, $[a,a^\dagger]=\1$, which look like $[a,b]=\1$, with $b\neq a^\dagger$. We have shown that, under suitable assumptions, two families of biorthogonal bases of the Hilbert space $\Hil$ on which $a$ and $b$ act can be constructed and that these sets may, or may not, be Riesz bases, see Section II. In particular, this characteristic is interesting since, as we have discussed in \cite{bag1}, any Riesz basis by itself produces non trivial examples of pseudo-bosons. In \cite{bag3} we have shown how to construct examples of pseudo-bosons by generalizing  standard techniques used in supersymmetric quantum mechanics, while in \cite{bag4} we have analyzed some mathematical aspect of our construction.  Here we continue our analysis more on a physical side, discussing in details the pseudo-bosonic structure for the extended quantum harmonic oscillator (EQHO), \cite{dapro}, and for the Swanson hamiltonian, \cite{dapro,swan,jon}.

 The paper is organized as follows: in order to keep the paper self-contained, in the next section we quickly review the general framework of pseudo-bosons.  In Section III we construct the biorthogonal bases for the EQHO. In Section IV we discuss the Swanson hamiltonian and its pseudo-bosonic excitations. Section V contains our conclusions.

\section{The general settings}

In this section we will review the general framework originally introduced in \cite{bag1} and further developed in \cite{bag2,bag3,bag4}.

Let $\Hil$ be a given Hilbert space with scalar product
$\left<.,.\right>$ and related norm $\|.\|$. We introduce a
pair of operators $a$ and $b$ acting on $\Hil$ and
satisfying the following commutation rules \be [a,b]=\1.
\label{21} \en  Of course, this collapses to the CCR's  if $b=a^\dagger$. It is well known
that $a$ and/or $b$ are unbounded operators, so they cannot be
defined in all of $\Hil$. Following \cite{bag1} we consider the
following

\vspace{2mm}

{\bf Assumption 1.} there exists a non-zero
$\varphi_{0}\in\Hil$ such that $a\varphi_{0}=0$,
and $\varphi_{0}\in D^\infty(b)$.

{\bf Assumption 2.} there exists a non-zero $\Psi_{0}\in\Hil$
such that $b^\dagger\Psi_{0}=0$,  and $\Psi_{0}\in
D^\infty(a^\dagger)$.

\vspace{2mm}

Here $D^\infty(b)$ is the domain of all the powers of $b$: $D^\infty(b)=\cap_{k\geq0} D(b^k)$, where $D(b^k)$ is the domain of $b^k$. Analogously $D^\infty(a^\dagger)$ is the domain of all the powers of $a^\dagger$. Under these assumptions we can introduce the following vectors of
$\Hil$: \be
\varphi_{n}=\frac{1}{\sqrt{n!}}\,b^n\,\varphi_{0}
\quad \mbox{ and }\quad
\Psi_{n}=\frac{1}{\sqrt{n!}}(a^\dagger)^n\Psi_{0,},
\quad n\geq 0. \label{22}\en Let us now define the unbounded
operators $N:=b\,a$ and $\N:=N^\dagger=a^\dagger
b^\dagger$.  It is possible to check that
$\varphi_{n}$ belongs to the domain of $N$, $D(N)$, and
$\Psi_{n}\in D(\N)$, for all $n\geq0$. Moreover,
\be N\varphi_{n}=n\varphi_{n}, \quad
  \N\Psi_{n}=n\Psi_{n}. \label{23}\en

Under the above assumptions it is  easy to check that
$\left<\Psi_{n},\varphi_{m}\right>=\delta_{n,m}\left<\Psi_{0},\varphi_{0}\right>$
for all $n, m\geq0$, which, if we chose the normalization of
$\Psi_{0}$ and $\varphi_{0}$ such that
$\left<\Psi_{0},\varphi_{0}\right>=1$, becomes \be
\left<\Psi_{n},\varphi_{m}\right>=\delta_{n,m},
\quad \forall n,m\geq0. \label{27}\en This means that the sets
$\F_\Psi=\{\Psi_{n},\,n\geq0\}$ and
$\F_\varphi=\{\varphi_{n},\,n\geq0\}$ are biorthogonal and,
because of this, the vectors of each set are linearly independent.
If we now call $\D_\varphi$ and $\D_\Psi$ respectively the linear
span of  $\F_\varphi$ and $\F_\Psi$, and $\Hil_\varphi$ and
$\Hil_\Psi$ their closures, then \be f=\sum_{n=0}^\infty
\left<\Psi_{n},f\right>\,\varphi_{n}, \quad \forall
f\in\Hil_\varphi,\qquad  h=\sum_{n=0}^\infty
\left<\varphi_{n},h\right>\,\Psi_{n}, \quad \forall
h\in\Hil_\Psi. \label{210}\en What is not in general ensured is
that all these Hilbert spaces do coincide, i.e. that
$\Hil_\varphi=\Hil_\Psi=\Hil$. Indeed, we can only state that
$\Hil_\varphi\subseteq\Hil$ and $\Hil_\Psi\subseteq\Hil$. However,
motivated by the examples already discussed in the literature and
by the results in Sections III and IV,  we consider

\vspace{2mm}

{\bf Assumption 3.} The above Hilbert spaces all coincide:
$\Hil_\varphi=\Hil_\Psi=\Hil$,

\vspace{2mm}

\noindent
which was introduced in \cite{bag1}. This means, in particular,
that both $\F_\varphi$ and $\F_\Psi$ are bases of $\Hil$. Let us
now introduce the operators $S_\varphi$ and $S_\Psi$ via their
action respectively on  $\F_\Psi$ and $\F_\varphi$: \be
S_\varphi\Psi_{n}=\varphi_{n},\qquad
S_\Psi\varphi_{n}=\Psi_{n}, \label{213}\en for all $n\geq0$, which also implies that
$\Psi_{n}=(S_\Psi\,S_\varphi)\Psi_{n}$ and
$\varphi_{n}=(S_\varphi \,S_\Psi)\varphi_{n}$, for all
$n\geq0$. Hence \be S_\Psi\,S_\varphi=S_\varphi\,S_\Psi=\1 \quad
\Rightarrow \quad S_\Psi=S_\varphi^{-1}. \label{214}\en In other
words, both $S_\Psi$ and $S_\varphi$ are invertible and one is the
inverse of the other. Furthermore, we can also check that they are
both well defined and symmetric, \cite{bag1}. Moreover, if $\left<\Psi_0,\varphi_o\right>=1$,  it is possible to write  these operators in a
bra-ket language as \be S_\varphi=\sum_{n=0}^\infty
|\varphi_{n}><\varphi_{n}|,\qquad S_\Psi=\sum_{n=0}^\infty
|\Psi_{n}><\Psi_{n}|. \label{212}\en These expressions are
only formal, at this stage, since the series could be not converging and the operators $S_\varphi$ and $S_\Psi$ could result unbounded.
This aspect was widely discussed in \cite{bag1}, where the role
of Riesz bases in relation with the boundedness of $S_\varphi$ and
$S_\Psi$ has been discussed in detail. For completeness' sake we recall that, given an orthonormal (o.n.) basis $\G=\{g_n,\,n\geq0\}$ of $\Hil$, and a bounded operator $X$ with bounded inverse, $\G^{(X)}:=\{g_n^{(X)}:=Xg_n,\,n\geq0\}$ is a Riesz basis of $\Hil$, \cite{you}. In particular, we have shown that $S_\varphi$ and $S_\Psi$ are bounded if and only if
$\F_\Psi$ and $\F_\varphi$ are Riesz bases. In this case we will call our excitations {\em regular pseudo-bosons}, otherwise we will just call them {\em pseudo-bosons}.
We will come back on this
aspect in the following sections.

\vspace{2mm}

We end this short review by recalling that these
 pseudo-bosons give rise to an interesting
intertwining relation among non self-adjoint operators, see
\cite{bag1}-\cite{bag4} and references therein. In particular it is easy to
check that \be S_\Psi\,N=\N S_\Psi \quad \mbox{ and }\quad
N \,S_\varphi=S_\varphi\,\N. \label{219}\en This is
related to the fact that the spectra of $N$ and $\N$
coincide, and that their eigenvectors are related by the operators
$S_\varphi$ and $S_\Psi$, in agreement with the literature on
intertwining operators, \cite{intop,bag5}.

\section{The extended quantum harmonic oscillator}

The hamiltonian of this model, introduced in \cite{dapro}, is the  non self-adjoint operator $H_\beta=\frac{\beta}{2}\left(p^2+x^2\right)+i\sqrt{2}\,p$, where $\beta$ is a positive parameter and $[x,p]=i$. This hamiltonian is not ${\mc PT}$-symmetric but satisfies $\PP H_\beta=H_\beta^\dagger \PP$, where $\PP$ and $\mc T$ are the parity and the time-reversal operators. In \cite{dapro} the {\em right} and {\em left} eigenvectors, as well as the spectrum of $H_\beta$, were found. Here we will see how this model can be discussed in the framework of Section II. In particular we will show that these right and left eigenvectors are nothing but the eigenvectors of $H_\beta$ and $H_\beta^\dagger$, which are automatically biorthogonal because of (\ref{27}). We will also prove that the two resulting sets are both complete in $\Lc^2(\Bbb{R})$, but they are not Riesz bases.

Introducing the standard bosonic operators $a=\frac{1}{\sqrt{2}}\left(x+\frac{d}{dx}\right)$, $a^\dagger=\frac{1}{\sqrt{2}}\left(x-\frac{d}{dx}\right)$, $[a,a^\dagger]=\1$, and the number operator $N=a^\dagger a$, we can write $H_\beta=\beta N+(a-a^\dagger)+\frac{\beta}{2}\,\1$ which, introducing now the operators
\be
A_\beta=a-\frac{1}{\beta}, \qquad B_\beta=a^\dagger+\frac{1}{\beta},
\label{31}\en
can be written as
\be
H_\beta=\beta\left(B_\beta A_\beta+\gamma_\beta\,\1\right),
\label{32}
\en
where $\gamma_\beta=\frac{2+\beta^2}{2\beta^2}$. It is clear that, for all $\beta>0$, $A_\beta^\dagger\neq B_\beta$ and that $[A_\beta, B_\beta]=\1$. Hence we have to do with pseudo-bosonic operators. This, of course, does not mean that the Assumptions of the previous section are necessarily satisfied; in \cite{bag1,bag3}  we have discussed examples of operators satisfying this commutation rule and for which some of the assumptions (or even all!) do not hold. However, this is not the case here: we will show that the Assumptions 1, 2 and 3 are indeed satisfied.

First of all, to check Assumption 1, we need to find a non zero vector $\varphi_0^{(\beta)}\in\Hil$ such that $A_\beta\varphi_0^{(\beta)}=0$ and $\varphi_0^{(\beta)}\in D^\infty(B_\beta)$. This is quite easy since condition $A_\beta\varphi_0^{(\beta)}=0$ can be rewritten as $a\,\varphi_0^{(\beta)}=\frac{1}{\beta}\,\varphi_0^{(\beta)}$. Hence, $\varphi_0^{(\beta)}$ is a standard coherent state with parameter $\frac{1}{\beta}$:
\be
\varphi_0^{(\beta)}=U(\beta^{-1})\varphi_0=e^{-1/2\beta^2}\,\sum_{k=0}^\infty\,\frac{\beta^{-k}}{\sqrt{k!}}\,\varphi_k,
\label{33}\en
where $\varphi_0$ is the vacuum of $a$, $a\varphi_0=0$, and $U(\beta^{-1})=e^{\frac{1}{\beta}(a^\dagger-a)}$ is the unitary (displacement) operator appearing in the theory of coherent states, \cite{gaz}. Moreover, the vectors $\varphi_k=\frac{(a^\dagger)^k}{\sqrt{k!}}\,\varphi_0$, $k\geq0$, in (\ref{33}) belong to the usual ortonormal (o.n.) basis $\F$ for an harmonic oscillator with hamiltonian $h=\omega\left(a^\dagger a+\frac{1}{2}\right)$. Of course, for all possible choices of $\beta$, we have $\|\varphi_0^{(\beta)}\|=\|\varphi_0\|=1$. In order to check whether $\varphi_0^{(\beta)}$ belongs to $D^\infty(B_\beta)$ we repeat similar calculations as in \cite{bag3}: since $U^{-1}(\beta^{-1})B_\beta U(\beta^{-1})=B_{\beta/2}$, and since $B_\beta^k\,\varphi_0^{(\beta)}=U(\beta^{-1})\left(U^{-1}(\beta^{-1})B_\beta U(\beta^{-1})\right)^k\varphi_0$, it is easy to check that $\|B_\beta^k\,\varphi_0^{(\beta)}\|\leq k!\,e^{2/\beta}$, $k\geq0$. Hence $\varphi_0^{(\beta)}$ belongs to the domain of all the powers of $B_\beta$, as required. This is crucial, since it implies that
\be
\varphi_n^{(\beta)}=\frac{1}{\sqrt{n!}}\,B_\beta^n \varphi_0^{(\beta)},
\label{34}\en
is well defined for all $n\geq 0$.

To check now the validity of Assumption 2 we first look for the solution of $B_\beta^\dagger\Psi_0^{(\beta)}=0$. This can be rewritten as $a\,\Psi_0^{(\beta)}=-\,\frac{1}{\beta}\,\Psi_0^{(\beta)}$, so that the solution is $\Psi_0^{(\beta)}=\varphi_0^{(-\beta)}=U(-\beta^{-1})\varphi_0=U^{-1}(\beta^{-1})\varphi_0$. Using now the equality
$U(\beta^{-1})A_\beta^\dagger U^{-1}(\beta^{-1})=A_{\beta/2}^\dagger$ we can also check that $\|(A_\beta^\dagger)^k\,\Psi_0^{(\beta)}\|\leq k!\,e^{2/\beta}$, $k\geq 0$. Hence
\be
\Psi_n^{(\beta)}=\frac{1}{\sqrt{n!}}\,(A_\beta^\dagger)^n \Psi_0^{(\beta)},
\label{35}\en
is also well defined for all $n\geq 0$. General reasons discussed in \cite{bag1} show that, calling $N_\beta=B_\beta A_\beta$ and $\N_\beta=N_\beta^\dagger=A_\beta^\dagger B_\beta^\dagger$, since
\be
N_\beta\,\varphi_n^{(\beta)}=n\,\varphi_n^{(\beta)}, \qquad \N_\beta \Psi_n^{(\beta)}= n\,\Psi_n^{(\beta)},
\label{36}\en
these vectors above are biorthogonal and, since $\left<\varphi_0^{(\beta)},\Psi_0^{(\beta)}\right>=e^{-2/\beta^2}$, the following holds:
\be
\left<\varphi_n^{(\beta)},\Psi_m^{(\beta)}\right>=\delta_{n,m} \,e^{-2/\beta^2}.
\label{37}\en

\vspace{2mm}

{\bf Remark:--} We could remove the factor $e^{-2/\beta^2}$ by changing the normalization of $\varphi_0^{(\beta)}$ and $\Psi_0^{(\beta)}$. We prefer to keep this normalization since it is standard for coherent states.

\vspace{2mm}

{\bf Remark:--} Since in the $x$-representation the coherent states of $a=\frac{1}{\sqrt{2}}\left(x+\frac{d}{dx}\right)$ with eigenvalues $\pm \frac{1}{\beta}$ look like $\frac{1}{\pi^{1/4}}\,e^{-\frac{1}{2}(x\mp \sqrt{2}/\beta)^2}$, it is possible to deduce the following expressions for the vectors introduced so far, which we now write introducing explicitly the dependence on $x$:
$$
\varphi_n^{(\beta)}(x)=\frac{1}{\pi^{1/4}\,\sqrt{2^n\,n!}}\,\left(x-\frac{d}{dx}+\frac{\sqrt{2}}{\beta}\right)^n\,e^{-\frac{1}{2}(x-\sqrt{2}/\beta)^2},
$$
and
$$
\Psi_n^{(\beta)}(x)=\frac{1}{\pi^{1/4}\,\sqrt{2^n\,n!}}\,\left(x-\frac{d}{dx}-\frac{\sqrt{2}}{\beta}\right)^n\,
e^{-\frac{1}{2}(x+\sqrt{2}/\beta)^2}.
$$
Not surprisingly, these functions coincide, except for some normalization constants, with the right and left eigenvectors in \cite{dapro}.

\vspace{2mm}

Let us now define the following sets of vectors: $\F_\varphi^{(\beta)}=\{\varphi_n^{(\beta)},\,n\geq 0\}$ and $\F_\Psi^{(\beta)}=\{\Psi_n^{(\beta)},\,n\geq 0\}$, their linear span $\D_\varphi^{(\beta)}$ and $\D_\Psi^{(\beta)}$, and the Hilbert spaces $\Hil_\varphi^{(\beta)}$ and $\Hil_\Psi^{(\beta)}$ obtained taking their closures.

Concerning Assumption 3, we will now prove that $\F_\varphi^{(\beta)}$ is complete in $\Hil$. To show this, we first observe that $U^{-1}(\beta^{-1})a^\dagger U(\beta^{-1})=B_\beta$, so that
$$
\varphi_n^{(\beta)}=\frac{1}{\sqrt{n!}}\,U^{-1}(\beta^{-1})(a^\dagger)U(2\beta^{-1})\varphi_0.
$$
Hence, we can check that a generic vector $f\in\Hil$ is orthogonal to all the $\varphi_n^{(\beta)}$'s if the following scalar products are all zero: $\left<U(-\beta^{-1})f,\left(a^\dagger+\frac{2}{\beta}\,\1\right)^n\varphi_0\right>=0$ for all $n\geq0$. This immediately implies that $\left<U(-\beta^{-1})f,(a^\dagger)^n\varphi_0\right>=0$ for all $n\geq0$, and therefore that $U(-\beta^{-1})f=0$, since $\F$ is an o.n. basis of $\Hil$. Hence $f=0$.

Similar techniques also show that $\F_\Psi^{(\beta)}$ is complete in $\Hil$. We conclude that Assumption 3 holds true.

To check whether $\F_\varphi^{(\beta)}$ and $\F_\Psi^{(\beta)}$ are Riesz bases or not it is convenient to introduce the following self-adjoint, unbounded and invertible operator: $V_\beta=e^{(a+a^\dagger)/\beta}$. This is useful because it turns out that
\be
A_\beta=V_\beta a V_\beta^{-1},\qquad B_\beta=V_\beta a^\dagger V_\beta^{-1}.
\label{38}\en
It should be stressed that these equalities, as well as many of those which will appear in the rest of this section, are only defined on the dense set $\F$, since they involve unbounded operators.

Formula (\ref{38}) implies that $H_\beta$ can be related to a self adjoint operator $h_\beta=\beta(a^\dagger a+\gamma_\beta\1)$ as $H_\beta=V_\beta h_\beta V_\beta^{-1}$ or, equivalently as
\be
H_\beta V_\beta=V_\beta h_\beta,
\label{39}\en
which shows that $V_\beta$ is an intertwining operator (IO) relating $h_\beta$ and $H_\beta$. Moreover, taking the adjoint of (\ref{39}), we get $V_\beta H_\beta^\dagger=h_\beta V_\beta$, so that $V_\beta$ is also an IO between $h_\beta$ and $H_\beta^\dagger$. This has well known consequences on the spectra of the three operators $h_\beta$, $H_\beta$ and $H_\beta^\dagger$ and on their eigenstates, \cite{intop}.

In particular we have $h_\beta\varphi_k=\epsilon_k^{(\beta)}\varphi_k$, where $\epsilon_k^{(\beta)}=\beta(k+\gamma_\beta)$, $\forall k\geq0$. Hence, calling $\Phi_k^{(\beta)}:=V_\beta\,\varphi_k$, which is clearly different from zero for all $k$ since $\ker(V_\beta)=\{0\}$, we have, using the intertwining equality (\ref{39}),
$$
H_\beta \Phi_k^{(\beta)}= H_\beta V_\beta\,\varphi_k=V_\beta h_\beta \varphi_k=\epsilon_k^{(\beta)}V_\beta  \varphi_k=\epsilon_k^{(\beta)}\Phi_k^{(\beta)}.
$$
Moreover, because of (\ref{36}), since $H_\beta\,\varphi_k^{(\beta)}=\beta(N_\beta+\gamma_\beta)\varphi_k^{(\beta)}=\epsilon_k^{(\beta)}\varphi_k^{(\beta)}$, and assuming that the eigenvalues $\epsilon_k^{(\beta)}$ are all non degenerate, it turns out that $\varphi_k^{(\beta)}=\alpha_k\,\Phi_k^{(\beta)}$ for all $k\geq0$, where $\alpha_k$ are simply complex constants. As a matter of fact we can further check that all these constants coincide: $\alpha_k=e^{-1/\beta^2}$, $k\geq0$,  so that, in conclusion,
\be
 \varphi_k^{(\beta)}=e^{-1/\beta^2}\,V_\beta \,\varphi_k,
 \label{310}\en
 for all $k\geq0$. Similar arguments can be repeated to analyze the eigensystem of $H_\beta^\dagger$: since $H_\beta^\dagger=V_\beta^{-1} h_\beta V_\beta$, if we put $\Upsilon_k^{(\beta)}=V_\beta^{-1}\varphi_k$ for all $k\geq0$, we get $H_\beta^\dagger\,\Upsilon_k^{(\beta)}=\epsilon_k^{(\beta)}\,\Upsilon_k^{(\beta)}$. Moreover, using (\ref{36}),
$H_\beta^\dagger\,\Psi_k^{(\beta)}=\beta(\N_\beta+\gamma_\beta)\Psi_k^{(\beta)}=\epsilon_k^{(\beta)}\,\Psi_k^{(\beta)}$. Therefore, because of the non degeneracy of $\epsilon_k^{(\beta)}$, we again deduce a proportionality between $\Upsilon_k^{(\beta)}$ and $\Psi_k^{(\beta)}$, which produces the equality
\be
 \Psi_k^{(\beta)}=e^{-1/\beta^2}\,V_\beta^{-1} \varphi_k,
 \label{310b}\en
for all $k\geq0$. This equation, together with (\ref{310}), also implies that $\Psi_k^{(\beta)}=V_\beta^{-2} \varphi_k^{(\beta)}$, $\forall k\geq0$. Therefore, recalling (\ref{213}), we recover the explicit expressions for the operators $S_\Psi^{(\beta)}$ and $S_\varphi^{(\beta)}$: $S_\Psi^{(\beta)}=V_\beta^{-2}$ and, consequently,  $S_\varphi^{(\beta)}=V_\beta^{2}$. This is in agreement with the following computations, extending formulas (\ref{212}) to the present situation where $\left<\varphi_0^{(\beta)},\Psi_0^{(\beta)}\right>=e^{-2/\beta^2}\neq 1$:
$$
\frac{1}{\left<\varphi_0^{(\beta)},\Psi_0^{(\beta)}\right>}\,\sum_{n=0}^\infty\,\left|\varphi_n^{(\beta)}\left>
\right<\Psi_n^{(\beta)}\right|=V_\beta\left(\sum_{n=0}^\infty\,\left|\varphi_n\left>
\right<\varphi_n\right|\right)V_\beta^\dagger=V_\beta^2=S_\varphi^{(\beta)},
$$
$$
\frac{1}{\left<\Psi_0^{(\beta)},\varphi_0^{(\beta)}\right>}\,\sum_{n=0}^\infty\,\left|\Psi_n^{(\beta)}\left>
\right<\varphi_n^{(\beta)}\right|=V_\beta^{-1}\left(\sum_{n=0}^\infty\,\left|\varphi_n\left>
\right<\varphi_n\right|\right)(V_\beta^{-1})^\dagger=V_\beta^{-2}=S_\Psi^{(\beta)},
$$
where we have used the closure relation for $\F$: $\sum_{k=0}^\infty\,\left|\varphi_n\right>\left<\varphi_n\right|=\1$. We also have
$$
\frac{1}{\left<\Psi_0^{(\beta)},\varphi_0^{(\beta)}\right>}\,\sum_{n=0}^\infty\,\left|\varphi_n^{(\beta)}\left>
\right<\Psi_n^{(\beta)}\right|=V_\beta\left(\sum_{n=0}^\infty\,\left|\varphi_n\left>
\right<\varphi_n\right|\right) V_\beta^{-1}=\1.
$$
Going back to the nature of the sets $\F_\varphi^{(\beta)}$ and $\F_\Psi^{(\beta)}$, since their vectors are obtained by the o.n. basis $\F$ via the action of the two unbounded operators $V_\beta$ and $V_\beta^{-1}$, they are not Riesz bases. Hence, Assumption 4 in \cite{bag1} is not satisfied: we have pseudo-bosons which are not regular.

\section{The Swanson hamiltonian}

The starting point is the following non self-adjoint hamiltonian, \cite{dapro}:
$$
H_\theta=\frac{1}{2}\left(p^2+x^2\right)-\frac{i}{2}\,\tan(2\theta)\left(p^2-x^2\right),
$$
where $\theta$ is a real parameter taking value in $\left(-\frac{\pi}{4},\frac{\pi}{4}\right)\setminus\{0\}=:I$. It is clear that $H_\theta^\dagger=H_{-\theta}\neq H_\theta$, for all $\theta\in I$. As usual, $[x,p]=i\1$. Introducing the annihilation and creation operators $a$ and $a^\dagger$ we write
$$
H_\theta=N+\frac{i}{2}\,\tan(2\theta)\left(a^2+(a^\dagger)^2\right)+\frac{1}{2}\,\1,
$$
where $N=a^\dagger a$. This hamiltonian can be rewritten  by introducing the operators
\be
\left\{
\begin{array}{ll}
A_\theta=\cos(\theta)\,a+i\sin(\theta)\,a^\dagger,  \\
B_\theta=\cos(\theta)\,a^\dagger+i\sin(\theta)\,a,
\end{array}
\right.
\label{41}\en
as
\be
H_\theta=\omega_\theta\left(B_\theta\,A_\theta+\frac{1}{2}\1\right),
\label{42}\en
where $\omega_\theta=\frac{1}{\cos(2\theta)}$ is well defined since $\cos(2\theta)\neq0$ for all $\theta\in I$. It is clear that $A_\theta^\dagger\neq B_\theta$ and that $[A_\theta,B_\theta]=\1$. Hence we can try to see if it is possible to construct pseudo-bosons out of this hamiltonian (i.e. if Assumptions 1, 2 and 3 hold), or, even more, if regular pseudo-bosons arise (i.e. if also Assumption 4 of \cite{bag1} is satisfied).
It will be convenient to rewrite (\ref{41}) by using the coordinate expressions for $a$ and $a^\dagger$:
\be
\left\{
\begin{array}{ll}
A_\theta=\frac{1}{\sqrt{2}}\left(e^{i\theta}x+e^{-i\theta}\,\frac{d}{dx}\right),  \\
B_\theta=\frac{1}{\sqrt{2}}\left(e^{i\theta}x-e^{-i\theta}\,\frac{d}{dx}\right).
\end{array}
\right.
\label{43}\en
We are now ready to check the validity of Assumptions 1 and 2. To begin with, we consider an {\em abstract} point of view: $\varphi_0^{(\theta)}$ satisfies $A_\theta\varphi_0^{(\theta)}=0$ if and only if $a\varphi_0^{(\theta)}=-i\tan(\theta)\,a^\dagger\,\varphi_0^{(\theta)}$. Expanding  $\varphi_0^{(\theta)}$ in both sides of this equality in terms of the o.n. basis $\F=\{\varphi_n,\,n\geq0\}$ introduced in the previous section, $\varphi_0^{(\theta)}=\sum_{n=0}^\infty\,c_n\varphi_n$, and taking the scalar product of the resulting equation with $\varphi_j$, $j=0,1,2,\ldots$, we deduce that $c_{2n+1}=0$ and $c_{2n}=(-i\tan(\theta))^n\,\sqrt{\frac{(2n-1)!!}{(2n)!!}}\,c_0$, $\forall n\geq0$. Here $(2n)!!=2n\cdot2(n-1)\cdot2(n-2)\cdots 4\cdot 2$ and $(2n+1)!!=(2n+1)\cdot(2n-1)\cdot(2n-3)\cdots 5\cdot 3\cdot 1$. Therefore we get
$$
\varphi_0^{(\theta)}=c_0\,\sum_{n=0}^\infty\,(-i\tan(\theta))^n\,\sqrt{\frac{(2n-1)!!}{(2n)!!}}\,\varphi_{2n}.
$$
This vector belongs to $\Hil$ for all $\theta\in I$. Indeed we have $\|\varphi_0^{(\theta)}\|^2=|c_0|^2\,\sum_{n=0}^\infty\,(\tan(\theta))^{2n}\,\frac{(2n-1)!!}{(2n)!!}$, which is always convergent for all $\theta\in I$.

Similar computations can be repeated to find the vector $\Psi_0^{(\theta)}$ satisfying $B_\theta^\dagger\Psi_0^{(\theta)}=0$. In this case we find
$$
\Psi_0^{(\theta)}=d_0\,\sum_{n=0}^\infty\,(i\tan(\theta))^n\,\sqrt{\frac{(2n-1)!!}{(2n)!!}}\,\varphi_{2n},
$$
and $\|\Psi_0^{(\theta)}\|^2=|d_0|^2\,\sum_{n=0}^\infty\,(\tan(\theta))^{2n}\,\frac{(2n-1)!!}{(2n)!!}$. To conclude that Assumptions 1 and 2 are verified, we still have to verify that the vectors we have found here belong to $D^\infty(B_\theta)$ and $D^\infty(A_\theta^\dagger)$ respectively. This would produce a rather long computation, which can be made much simpler if we work directly in the coordinate representation, i.e. considering the expressions (\ref{43}) of our pseudo-bosonic operators. From now on, this will be our point of view.

Equation $A_\theta\varphi_0^{(\theta)}=0$ becomes $\left(e^{i\theta}x+e^{-i\theta}\frac{d}{dx}\right)\varphi_0^{(\theta)}(x)=0$ whose solution is
\be
\varphi_0^{(\theta)}(x)=N_1 \exp\left\{-\frac{1}{2}\,e^{2i\theta}\,x^2\right\},
\label{44}\en
where $N_1$ is a normalization constant. Analogously, $B_\theta^\dagger\Psi_0^{(\theta)}=0$ becomes  $\left(e^{-i\theta}x+e^{i\theta}\frac{d}{dx}\right)\Psi_0^{(\theta)}(x)=0$, so that
\be
\Psi_0^{(\theta)}(x)=N_2 \exp\left\{-\frac{1}{2}\,e^{-2i\theta}\,x^2\right\},
\label{45}\en
where, again, $N_2$ is a normalization constant. Notice that, since $\Re(e^{\pm 2i\theta})=\cos(2\theta)>0$ for all $\theta\in I$, both $\varphi_0^{(\theta)}(x)$ and $\Psi_0^{(\theta)}(x)$ belong to $\Lc^2({\Bbb R})$, which is the Hilbert space $\Hil$ of the theory.

Defining now the vectors $\varphi_n^{(\theta)}(x)$ and $\Psi_n^{(\theta)}(x)$ as in (\ref{22}), we find the following interesting result:

\be
\left\{
\begin{array}{ll}
\varphi_n^{(\theta)}(x)=\frac{1}{\sqrt{n!}}\,B_\theta^n\,\varphi_0^{(\theta)}(x)=\frac{N_1}{\sqrt{2^n\,n!}}
\,H_n\left(e^{i\theta}x\right)\,\exp\left\{-\frac{1}{2}\,e^{2i\theta}\,x^2\right\},  \\
\Psi_n^{(\theta)}(x)=\frac{1}{\sqrt{n!}}\,(A_\theta^\dagger)^n\,\Psi_0^{(\theta)}(x)=\frac{N_2}{\sqrt{2^n\,n!}}
\,H_n\left(e^{-i\theta}x\right)\,\exp\left\{-\frac{1}{2}\,e^{-2i\theta}\,x^2\right\},
\end{array}
\right.
\label{46}\en
where $H_n(x)$ is the n-th Hermite polynomial. These equalities can be proved  using induction on $n$ and  the following identity of Hermite polynomials: $H_{n+1}(x)=2xH_n(x)-2nH_{n-1}(x)$. The norm of the vectors in (\ref{46}) can be computed using the following formula:
$$
\int_0^\infty e^{-px^2}\,H_n(bx)\,H_n(cx)\,dx=\frac{2^{n-1}\,n!\,\sqrt{\pi}}{p^{(n+1)/2}}\,(b^2+c^2-p)^{n/2}\,
P_n\left(\frac{bc}{\sqrt{p(b^2+c^2-p)}}\right)
$$
which holds for all $p$ with positive real part, \cite{prud}. Here $P_n$ is the n-th Legendre polynomial. Therefore we find
$$
\|\varphi_n^{(\theta)}\|^2=|N_1|^2\,\cos\left(\frac{\pi}{\cos(2\theta)}\right)\,
P_n\left(\frac{1}{\cos(2\theta)}\right)
$$
and
$$
\|\Psi_n^{(\theta)}\|^2=|N_2|^2\,\cos\left(\frac{\pi}{\cos(2\theta)}\right)\,
P_n\left(\frac{1}{\cos(2\theta)}\right),
$$
which are both well defined (even if the argument of $P_n$ does not belong to the interval $[-1,1]$), for all fixed $n$. Hence Assumptions 1 and 2 are necessarily satisfied. This was already clear from (\ref{46}), since the product of a polynomial of any order times a gaussian is square integrable. What is not clear at this stage is whether the sets $\F_\varphi^{(\theta)}=\{\varphi_n^{(\theta)}(x),\,n\geq0\}$ and $\F_\Psi^{(\theta)}=\{\Psi_n^{(\theta)}(x),\,n\geq0\}$ are (i) complete in $\Lc^2(\Bbb{R})$; (ii) Riesz bases.

To answer to the first question we use the same general idea adopted in \cite{bag3}, which was based on a result discussed in \cite{kolfom}: if $\rho(x)$ is a Lebesgue-measurable function which is different from zero almost everywhere (a.e.) in $\Bbb R$ and if there exist two positive constants $\delta, C$ such that $|\rho(x)|\leq C\,e^{-\delta|x|}$ a.e. in $\Bbb R$, then the set $\left\{x^n\,\rho(x)\right\}$ is complete in $\Lc^2(\Bbb{R})$.

Having this in mind, first of all we notice that $\F_\varphi^{(\theta)}$ is complete in $\Lc^2(\Bbb{R})$ if and only if the set $\F_\pi^{(\theta)}:=\{\pi_n^{(\theta)}(x)=x^n\,
\exp\left\{-\frac{1}{2}\,e^{2i\theta}\,x^2\right\},\,n\geq0\}$ is complete in $\Lc^2(\Bbb{R})$. Hence, because of the above cited result and since $\exp\left\{-\frac{1}{2}\,e^{2i\theta}\,x^2\right\}$ satisfies for our values of $\theta$ the conditions required to $\rho(x)$, then $\F_\pi^{(\theta)}$ is complete and, as a consequence,
$\F_\varphi^{(\theta)}$ is complete in $\Lc^2(\Bbb{R})$. The same conclusion can be deduced for the set $\F_\Psi^{(\theta)}$, which is therefore also complete in $\Lc^2(\Bbb{R})$. Therefore, Assumption 3 is satisfied: $\Hil_\varphi=\Hil_\Psi=\Hil$.

Let us now go back to the biorthogonality of the two sets $\F_\varphi^{(\theta)}$ and $\F_\Psi^{(\theta)}$. Condition $\left<\varphi_0^{(\theta)},\Psi_0^{(\theta)}\right>=1$ is ensured by requiring that $\overline{N_1}\,N_2=\frac{e^{-i\theta}}{\sqrt{\pi}}$. Hence, with this choice, we know that $\left<\varphi_n^{(\theta)},\Psi_m^{(\theta)}\right>=\delta_{n,m}$ which can be written explicitly as
$$
\int_{\Bbb{R}}H_n\left(e^{-i\theta}x\right)H_m\left(e^{-i\theta}x\right)e^{-e^{-2i\theta}x^2}\,dx=\delta_{n,m}\,\sqrt{2^{n+m}\,\pi\,n!\,m!},
$$
which is a non trivial integral which can also be found in \cite{prud}.

To understand whether our biorthogonal sets are Riesz bases or not we will proceed as in the previous section, by introducing a certain unbounded operator and showing for instance that this operator maps an o.n. basis of $\Hil$ into the set $\F_\varphi^{(\theta)}$. This would imply that $\F_\varphi^{(\theta)}$ cannot be a Riesz basis.

Let us introduce the following unbounded, self-adjoint and invertible operator $T_\theta=e^{i\frac{\theta}{2}(a^2-{a^\dagger}^2)}$. Then we have
\be
A_\theta=T_\theta a T_\theta^{-1},\qquad B_\theta = T_\theta a^\dagger T_\theta^{-1}.
\label{47}\en
This implies that $H_\theta=T_\theta h_\theta T_\theta^{-1}$, where $h_\theta=\omega_\theta\left(a^\dagger a+\frac{1}{2}\1\right)$. Hence, as in the previous model, $T_\theta$ is an IO:
\be
H_\theta T_\theta=T_\theta h_\theta,\qquad  T_\theta H_\theta^\dagger= h_\theta T_\theta
\label{48}\en
The same arguments discussed in Section III show that, if the eigenvalues $\omega_n^{(\theta)}=\omega(n+1/2)$ are non degenerate, then a single complex constant $\alpha$ must exist such that
\be
\varphi_n^{(\theta)}=\alpha\, T_\theta\, \varphi_n, \quad \mbox{ and }\quad \Psi_n^{(\theta)}=\frac{1}{\overline\alpha}\, T_\theta^{-1}\, \varphi_n
\label{49}\en
These equalities show, in particular, that neither $\F_\varphi^{(\theta)}$ nor $\F_\Psi^{(\theta)}$ are Riesz bases. Also, we deduce that $S_\varphi^{(\theta)}=|\alpha|^2\,T_\theta^2$ and $S_\Psi^{(\theta)}=|\alpha|^{-2}\,T_\theta^{-2}$. This is in agreement with the following (formal) computations:
$$
\sum_{n=0}^\infty\,\left|\varphi_n^{(\theta)}\left>
\right<\varphi_n^{(\theta)}\right|=\alpha\,T_\theta\left(\sum_{n=0}^\infty\,\left|\varphi_n\left>
\right<\varphi_n\right|\right)(\alpha T_\theta)^\dagger=|\alpha|^2 T_\theta^2=S_\varphi^{(\beta)},
$$
as well as
$$
\sum_{n=0}^\infty\,\left|\Psi_n^{(\theta)}\left>
\right<\Psi_n^{(\theta)}\right|=\frac{1}{\overline\alpha}\,T_\theta^{-1}\left(\sum_{n=0}^\infty\,\left|\varphi_n\left>
\right<\varphi_n\right|\right)\left(\frac{1}{\overline\alpha}\,T_\theta^{-1}\right)^\dagger=\frac{1}{|\alpha|^2} T_\theta^{-2}=S_\Psi^{(\beta)}.
$$
The resolution of the identity looks like
$$
\sum_{n=0}^\infty\,\left|\varphi_n^{(\theta)}\left>
\right<\Psi_n^{(\theta)}\right|=\alpha\,T_\theta\left(\sum_{n=0}^\infty\,\left|\varphi_n\left>
\right<\varphi_n\right|\right)\frac{1}{\alpha}\,T_\theta^{-1}=\1.
$$
\vspace{3mm}

Before closing the section it is interesting to notice that these results could be slightly generalized by  reversing the point of view we have considered so far: up to now we have considered non self-adjoint hamiltonians which  can be written essentially as pseudo-bosonic number operators. Then we have discovered that an unbounded map exists which transforms the pseudo-bosonic operators into {\em standard bosons}, see (\ref{38}) and (\ref{47}). Now we start considering the unbounded  map $T_{\alpha,\beta}=e^{\alpha a^2+\beta {a^\dagger}^2}$ depending on two real parameters $\alpha$ and $\beta$ and we define two operators $A_{\alpha,\beta}=T_{\alpha,\beta}\,a\,T_{\alpha,\beta}^{-1}$ and $B_{\alpha,\beta}=T_{\alpha,\beta}\,a^\dagger\,T_{\alpha,\beta}^{-1}$. It is possible to deduce that $A_{\alpha,\beta}=a\,\cos(\sqrt{4\alpha\beta})-\sqrt{\frac{\beta}{\alpha}}\,\sin(\sqrt{4\alpha\beta})\,a^\dagger$ and
$B_{\alpha,\beta}=a^\dagger\,\cos(\sqrt{4\alpha\beta})+\sqrt{\frac{\alpha}{\beta}}\,\sin(\sqrt{4\alpha\beta})\,a$, which is clearly different from $A_{\alpha,\beta}^\dagger$, in general. Moreover, $[A_{\alpha,\beta},B_{\alpha,\beta}]=\1$. Hence, we have easily constructed a pseudo-bosonic commutation rule.
However, there is not a big difference between these operators and the operators $A_\theta$, $B_\theta$ considered above in this section. For this reason we will not construct the biorthogonal sets arising from $A_{\alpha,\beta}$ and $B_{\alpha,\beta}$: the main steps will not differ significantly from those already considered here.

\section{Conclusions}

In this note we have shown that some non self-adjoint hamiltonians related to relevant quantum models can be analyzed within the framework of pseudo-bosons. It is also shown that these models give not rise to regular pseudo-bosons, since the biorthogonal bases they produce are not Riesz bases. This result, together with other models considered so far, suggests that pseudo-bosons may have a real physical interpretation while regular pseudo-bosons are mainly mathematical objects. This suggests that unbounded operators play a crucial role in the analysis of physical pseudo-bosons, which, in a sense, is in agreement with all the literature on the subject: most quantum mechanical systems (but for those living in finite dimensional Hilbert spaces, for which the operators are just matrices whose spectra are automatically bounded) depend on unbounded operators! Regularity looks more like a mathematical (rather than physical) requirement. Consequently, the role of Riesz bases appear more interesting from a mathematical rather than from a physical point of view.

A natural question to consider is now the following: how much of this general structure can be extended to generic non-hermitian hamiltonians? This is work in progress.

\section*{Acknowledgements}

It is a pleasure to thank Prof. Trapani for many interesting discussions (as always!!).  The author also acknowledges financial support by the Murst, within the  project {\em Problemi
Matematici Non Lineari di Propagazione e Stabilit\`a nei Modelli
del Continuo}, coordinated by Prof. T. Ruggeri.


\begin{thebibliography}{99}

\bibitem{bag1} F. Bagarello {\em Pseudo-bosons, Riesz bases and coherent states}, J. Math. Phys., {\bf 50}, DOI:10.1063/1.3300804, 023531 (2010) (10pg)



\bibitem{bag2} F. Bagarello, F. Calabrese {\em Pseudo-bosons arising from Riesz bases}, Bollettino del Dipartimento di Metodi e Modelli Matematici,  {\bf 2}, 15-26, (2010)

\bibitem{bag3} F. Bagarello {\em Construction of pseudo-bosons systems},  J. Math. Phys., {\bf 51}, DOI:10.1063/1.3300804, 023531 (2010) (10pg)

\bibitem{bag4} F. Bagarello {\em Mathematical aspects of intertwining
operators: the role of Riesz bases},   J. Phys. A, DOI:10.1088/1751-8113/43/17/175203, {\bf 43},  175203 (2010) (12pp)


\bibitem{dapro} J. da Provid$\hat e$ncia, N. Bebiano, J.P. da Provid$\hat e$ncia, {\em Non hermitian operators with real spectrum in quantum mechanics}, arXiv:0909.3054 [quant-ph]

\bibitem{swan} M. S. Swanson, {Transition elements for a non-Hermitian quadratic Hamiltonian},
 J. Math. Phys., {\bf 45}, 585, (2004)

\bibitem{jon} H.F. Jones, {On pseudo-hermitian hamiltonians and their hermitian counterparts},
 J. Phys. A, {\bf 38}, 1741, (2005)

\bibitem{you} Young R., {\em An introduction to nonharmonic Fourier series}, Academic Pree, New York, (1980); Christensen O., {\em An Introduction to Frames and Riesz Bases}, Birkh\"auser, Boston, (2003)


\bibitem{intop} Kuru S., Tegmen A., Vercin A., {\em Intertwined isospectral potentials in an arbitrary dimension},
J. Math. Phys, {\bf 42}, No. 8,
3344-3360, (2001); Kuru S., Demircioglu B., Onder M., Vercin A., {\em Two families of superintegrable and isospectral potentials in two dimensions},
J. Math. Phys, {\bf 43}, No. 5,
2133-2150, (2002); Samani K. A., Zarei M., {\em Intertwined hamiltonians in two-dimensional curved spaces}, Ann. of Phys., {\bf 316}, 466-482, (2005).


\bibitem{bag5} F. Bagarello {\em Extended SUSY quantum mechanics, intertwining operators and coherent states},
  Phys. Lett. A, DOI: 10.1016/ j.physleta. 2008.08.047 (2008), F. Bagarello {\em Vector coherent states and intertwining operators};
   J. Phys. A., DOI:10.1088/1751-8113/42/7/075302, (2009);
F. Bagarello, {\em Intertwining operators between different Hilbert spaces: connection with frames},  J. Math. Phys., DOI: 10.1063/1.3094758, {\bf 50}, 043509 (2009) (13pp)

\bibitem{gaz} J.-P. Gazeau, {\em Coherent states in quantum physics}, Wiley-VCH, Weinheim (2009)



\bibitem{prud} A. P. Prudnikov, Yu. A. Bryehkov, O.I. Marichev, {\em Integrals and series}, vol. 2, Special functions, Opa, Amsterdam, (1986)

\bibitem{kolfom} A. Kolmogorov and S. Fomine, {\em El\'ements de la th\'eorie des fonctions et de l'analyse fonctionnelle}, Mir (1973)


\end{thebibliography}
\end{document}